\documentclass{article}
\usepackage{graphicx}
\usepackage{amsmath, amssymb}
\usepackage{bm}
\usepackage{float}
\usepackage{color}
\usepackage[boxed]{algorithm2e}
\usepackage{upgreek}
\usepackage{multirow}
\usepackage{braket}
\usepackage{soul}
\usepackage[margin=1in]{geometry}

\graphicspath{{figures/}}

\newcommand{\norm}[1]{\left\lVert#1\right\rVert}

\def\deltares{\delta_{r}}

\newcommand{\micron}{$\upmu$m}

\DeclareMathOperator*{\argmin}{arg\,min}

\newcommand{\vecx}{\boldsymbol{n}}
\newcommand{\vecy}{\boldsymbol{y}}

\newcommand{\photonfluence}{n_{\textup{ph}}}
\newcommand{\snr}{\mbox{SNR}}

\newcommand{\freqr}{u_{r}}
\newcommand{\freqri}{u_{r,i}}
\newcommand{\realpix}{\Delta_{r}} 

\begin{document}

\title{Near, far, wherever you are: simulations on the dose
  efficiency of holographic and ptychographic coherent imaging}
\author
{Ming Du$^{1}$, Dog{\u a} G{\"u}rsoy$^{2,3}$, and Chris Jacobsen$^{2,4,5,\ast}$ \\
\normalsize{$^{1}$Department of Materials Science, Northwestern
  University, Evanston, IL 60208, USA} \\
\normalsize{$^{2}$Advanced Photon Source, Argonne National Laboratory,
  Argonne, IL 60439, USA} \\
\normalsize{$^{3}$Department of Electrical Engineering and Computer Science, 
  Northwestern University, Evanston, IL 60208, USA} \\
\normalsize{$^{4}$Department of Physics and Astronomy, Northwestern
  University, Evanston, IL 60208, USA} \\
\normalsize{$^{5}$Chemistry of Life Processes Institute, Northwestern
  University, Evanston, IL 60208, USA} \\
\\
\normalsize{$^\ast$To whom correspondence should be addressed; E-mail:
  cjacobsen@anl.gov}
}

\date{}

\maketitle

\begin{abstract}
  Different studies in x-ray microscopy have arrived at conflicting
  conclusions about the dose efficiency of imaging modes involving the
  recording of intensity distributions in the near (Fresnel regime) or
  far (Fraunhofer regime) field downstream of a specimen.  We present
  here a numerical study on the dose efficiency of near-field
  holography (NFH), near-field ptychography (NFP),
  and far-field ptychography (FFP), where ptychography
    involves multiple overlapping finite-sized
  illumination positions.  Unlike what has been reported for
  coherent diffraction imaging (CDI), which involves
    recording a single far-field diffraction pattern, we find that all
    three methods offer similar image quality when using the same
    fluence on the specimen, with far-field ptychography offering
    slightly better spatial resolution and lower mean error. These
  results support the concept that (if the experiment and image
  reconstruction are done properly) the sample can be near, or far;
  wherever you are, photon fluence on the specimen sets one limit to
  spatial resolution.
\end{abstract}

\section{Introduction}

X-ray microscopy provides a unique combination of short wavelength
radiation (with the potential for nanoscale imaging), with high
penetration.  However, X rays ionize atoms, so radiation damage often
sets a limit on the achievable resolution, especially when studying
soft or biological materials \cite{sayre_science_1977,kirz_qrb_1995}.
This becomes quite important as one seeks finer spatial resolution
$\deltares$, since for isotropic objects there is a
tendancy \cite{sayre_ultramic_1977,howells_jesrp_2009} for the required
number of photons per area incident on the specimen
  (the fluence $\photonfluence$) to obtain an image with sufficient
signal-to-noise ratio to increase as
$\photonfluence \propto (\deltares)^{-4}$.  Since fluence leads
directly to the absorbed radiation dose $D$, it is important to use
low-fluence methods for high resolution imaging.

One of the methods for low-fluence and low-dose x-ray imaging is to
use phase contrast.  That is because
\cite{henke_adndt_1993,du_ultramic_2018} the phase shift imparted on
an x-ray wavefront scales like $\rho Z\lambda^{2}$, while beam
absorption scales like $\rho Z \lambda^{4}$, where $\rho$ is the
density, $Z$ is the atomic number, and $\lambda$ is the wavelength.
As a result, phase contrast often leads to reduced radiation dose for
the same feature detectability, especially at shorter wavelengths
\cite{schmahl_xrmtaiwan}.

While the phase of an x-ray wave cannot be measured directly, it can
be inferred by mixing with a reference wave so that phase changes are
encoded as intensity differences.  This can be done using the Zernike
method with x-ray zone plates \cite{schmahl_optik_1994}, or by using
beam propagation.  In near-field methods involving
  short propagation distances from the specimen to a detector, one or
a few Fresnel fringes can be interpreted using approaches such as the
transport of intensity \cite{paganin_jmic_2002}, while at intermediate
distances a large number of Fresnel fringes allow for in-line
holographic reconstruction \cite{baez_nature_1952,baez_josa_1952}
in an approach that is often referred to as near-field
  holography (NFH).  One can improve reconstruction fidelity in
  near-field holography by combining information from holograms
  recorded at multiple distances \cite{cloetens_apl_1999}, or from
  multiple lateral illumination shifts \cite{stockmar_scirep_2013}
  where the latter approach is referred to as near-field pytchography
  (NFP).  If instead the beam is allowed to propagate from the
  specimen to a detector at a distance that meets the far-field or
Fraunhofer condition, x-ray images of phase objects can be recovered
from coherent diffraction patterns with no wave mixing required
\cite{sayre_schlenker_1980}.  This can be done in a single
illumination approach now called coherent diffraction imaging or CDI
\cite{miao_nature_1999} where one uses finite support iterative phase
retrieval \cite{fienup_optlett_1978}. Alternatively, it can be done
using multiple finite-sized overlappping coherent illumination spots
in a method called far-field ptychography (FFP)
\cite{hoppe_aca1_1969,hoppe_aca3_1969}, where one again uses an
iterative phase retrieval algorithm \cite{faulkner_prl_2004} to obtain
an image with a spatial resolution much finer than the size of the
illumination spot \cite{rodenburg_prl_2007}.

Are there fundamental differences in photon exposure requirements
depending on whether one mixes the specimen wave with a reference to
get intensities, or measures the specimen wave diffraction intensities
alone?  One might think that by mixing a strong reference wave $R$
with a weak specimen wave $S$ one might have a multiplying effect due
to the net intensity recording being
$|R|^{2}+RS^{\dag}+R^{\dag}S+|S|^{2}$, and indeed it has been
suggested that near-field x-ray holography (NFH) might be an
especially dose-efficient imaging method
\cite{bartels_prl_2015,hagemann_jac_2017} though other simulation
studies by some of the same researchers have found more of a
dose-equivalence with far-field diffraction \cite{jahn_aca_2017}.  In
fact, quantum noise is still limited by the specimen wave, leading to
the following conclusion by Richard Henderson
\cite{henderson_qrb_1995} in the context of electron microscopy: ``It
can be shown that the intensity of a sharp diffraction spot containing
a certain number $N$ of diffracted quanta will be measured with the
same accuracy ($\sqrt{N}$) as would the amplitude (squared) of the
corresponding Fourier component in the bright field phase contrast
image that would result from interference of this scattered beam with
the unscattered beam \cite{henderson_ultramic_1992}. The diffraction
pattern, if recorded at high enough spatial resolution, would
therefore contain all the intensity information on Fourier components
present in the image.''  This point is also addressed
  in Sec.~4.8.5 of \cite{jacobsen_2020}.  This leads us to expect
  that the reconstruction of a certain spatial frequency of the
object should be equally accurate for far-field diffraction
as it is for near-field phase contrast imaging,
provided both use the same fluence $\photonfluence$ on a specimen
pixel.

One could argue that the act of recovering phases from far-field
diffraction patterns can introduce extra noise.  Indeed, Henderson
followed the comments above \cite{henderson_qrb_1995} with this
statement: ``It [the diffraction pattern] would lack only the
information concerning the phases of the Fourier components of the
image which are of course lost.  Thus, for the same exposure,
holography should be equal to normal phase contrast in performance,
and diffraction methods inferior because of the loss of the
information on the phases of the Fourier components of the image.''
However, diffraction patterns \emph{are} affected by the phase of
Fourier components.  Consider the example of a transverse shift of one
subregion of a coherently illuminated object: the shift theorem of the
Fourier transform makes it clear that one would change the phase of
that subregion's contribution to a specific point in the entire
object's complex diffraction amplitude.  Therefore the intensity of
the diffraction pattern produced by the object would undergo some
redistribution (that is, the speckle pattern would change), showing
that diffraction methods do indeed involve the encoding of phase.
This is perhaps why a number of studies on iterative phase retrieval
methods have indicated that the phase retrieval process seems not to
add additional noise to the reconstructed image beyond that present in
the diffraction pattern itself
\cite{fienup_optlett_1978,williams_aca_2007,huang_optexp_2009,schropp_njp_2010,godard_optexp_2012}.

A slightly different approach to compare the signal to
  noise ratio for various imaging methods is to consider the strength
  of the signal scattered by a Gaussian-shaped feature characterized
  by a width $\sigma_{f}$, relative to the signal from the total
  illuminated area (the field of view or FOV)
  \cite{villanueva_optexp_2016}.  Unlike calculations that assume
  isotropic features and then calculate their contrast based on x-ray
  interaction properties
  \cite{sayre_ultramic_1977,howells_jesrp_2009,du_ultramic_2018}, this
  approach assumes that the feature scatters some number $N_{s}$ of
  photons for a given incident illumination ($N_{s}$ can be estimated
  \cite{shen_jsr_2004,schropp_njp_2010,villanueva_optexp_2016}).  This
  approach has been used to calculate a signal-to-noise ratio (SNR)
  for propagation-based phase contrast microscopy (PM; Eq.~8 of
  \cite{villanueva_optexp_2016}) of
  \begin{equation}
    \text{SNR}^{\text{PM}} \approx
    2\sqrt{N_S}\frac{4}{\sqrt{\pi}}B\frac{2\sigma_{f}}{\text{FOV}_{\text{PM}}}
    \label{eqn:snr_pm}
  \end{equation}
  where $B= p_s/f_w$ is the ratio of pixel size $p_{s}$ over source
  size $f_{w}$ ($B=1$ for coherent plane wave illumination from a
  distant source in near-field holography or NFH).  Analysis of
  coherent diffraction imaging (CDI) coherent diffraction imaging
  (CDI; Eq.~9 of \cite{villanueva_optexp_2016}) yields an expression
  of
  \begin{equation}
    \text{SNR}^{\text{CDI}} \approx
    \sqrt{N_S}\frac{2\sigma_{f}}{\mbox{FOV}_{\text{CDI}}}.
    \label{eqn:snr_cdi}
  \end{equation}
  The field of view (FOV) of CDI in Eq.~\ref{eqn:snr_cdi} can be
  reinterpreted as the probe size in far-field ptychography (FFP). For
  a Gaussian probe, a reasonable way to define the probe size would be
  to consider a sharp-edged disk concentric with the probe, and having
  the same height (1) as the probe's magnitude distribution.  For the
  disk to have the same integral area as the probe, its diameter needs
  to be $2\sqrt{2}$ times the Gaussian probe's standard deviation
  $\sigma_{f}$. With this assumption, and expressing the field of view
  (FOV) as (512 pixels) for NFH and ($2\sqrt{2)}$ pixels) for FFP, the
  SNR ratio between far-field ptychography (FFP) and near-field
  holography (NFH) becomes
  \begin{equation}
    \frac{\text{SNR}^{\text{FFP}}}{\text{SNR}^{\text{NFH}}}
    =
    \frac{\sqrt{\pi}}{8\beta}\frac{\text{FOV}_{\text{NFH}}}{\text{FOV}_{\text{FFP}}}
    = \frac{\sqrt{\pi}}{8\times 1}\times\frac{512}{2\sqrt{2}\, 6}
    \approx 6.7,
    \label{eqn:snr_ratio_ffp_nfh}
  \end{equation}
  suggesting that far-field ptychography (FFP) has a slight advantage
  over near-field holography.

In spite of Eq.~\ref{eqn:snr_ratio_ffp_nfh}, we hypothesize that the
fluence $\photonfluence$ on the object sets the main limit on
achievable resolution, rather than the use of near-field versus
far-field imaging methods (assuming both methods are implemented in a
way that allows a specific spatial resolution target to be reached).
This hypothesis is supported by a previous simulation study of binary
objects using propagation with different Fresnel numbers
\cite{jahn_aca_2017}. Excluding the contact regime (where one loses
sensitivity to phase contrast), this work concluded that near-field
and far-field imaging methods require essentially the same critical
photon fluence to reach the same level of reconstruction
error. Nevertheless, this analysis was carried out using small objects
with binary contrast and within rectangular supports, whereas we
examine below the same irregularly-sized objects with more continuous
contrast that were used in a different near-field/far-field comparison
\cite{hagemann_jac_2017}. In addition, both this binary object study
\cite{jahn_aca_2017} and other previous studies
\cite{huang_optexp_2009,hagemann_jac_2017} used single diffraction
patterns from finite-sized objects for far-field imaging.  The
reconstruction of complex objects from their single coherent
diffraction patterns is not always straightforward, as one needs
precise knowledge of the specimen's support $S$ (the subregion within
which the object is restricted to lie
\cite{fienup_josaa_1987,huang_optexp_2010}).  In addition, other
experimental limitations like the loss of a significant subset of
strong, low-spatial-frequency intensity values due to the presence of
beam stops can complicate object reconstruction
\cite{thibault_actaa_2006,huang_optexp_2010}.  These complications may
have played a role in the simulation study noted earlier that showed
that NFH yields superior images at the same fluence $\photonfluence$
when compared to using standard CDI as a far-field imaging method
\cite{hagemann_jac_2017}.

The problems noted above for standard CDI are greatly mitigated in
FFP, where the finite coherent illumination spot provides several
benefits. Ptychography allows one to accurately determine
the equivalent of a finite
support due not to the characteristics of the object, but instead due
to the characteristics of the limited-size probe function, which
can be recovered from the data.  Object subregions
that are present in the overlap between two probe positions provide a
sort of holographic reference between the two resulting diffraction
patterns \cite{bunk_ultramic_2008}. Finally, the spreading of the
unmodulated probe function in the far field (due to its finite extent
at the object's plane) helps distribute intensities out of the
central, zero-spatial-frequency pixel on the diffraction detector,
especially when the probe is a convergent beam provided by the focus
of a lens \cite{thibault_science_2008}.  Therefore while standard CDI
often shows imperfections in image reconstruction beyond those
provided by fluence, FFP can provide a method for a more robust
comparison between the fluence requirements of near-field versus
far-field coherent imaging methods.  It is for these reasons that we
have carried out a simulation study comparing NFH not against CDI, but
against FFP as a far-field imaging method.  We also
  include a comparison with near-field ptychography (NFP) as a method
  that combines near-field recording as in near-field holography
  (NFH), with multiple illumination positions as in far-field
  ptychography (FFP).

\section{Image reconstruction method}

In order to compare different imaging methods for non-binary objects,
we have chosen to use the same optimization-based reconstruction
method for the three imaging approaches, so as reveal only the
inherent differences between them.  The work of Hagemann and Salditt
\cite{hagemann_jac_2017} used the relaxed averaged alternating
reflections or RAAR algorithm \cite{luke_ip_2005} for reconstruction.
We have chosen to
make use of the same simulated object that they used (shown in Fig.~2(a) of
\cite{hagemann_jac_2017}).  However, in our case we have chosen to use a more
basic cost function minimization approach, in which one defines a
forward model for how incident illumination interacts with a present
guess of the object to produce a measurable intensity distribution,
after which one seeks adjust the object guess so as to minimize the
difference between the result of this forward model and the actual
measured intensity distribution (we refer to this difference as the
cost function $C$).  One can also include regularizers in this
approach as will be described below.  In order to efficiently minimize
the cost function $C$ for the three different imaging methods of
NFH, FFP, and NFP, we have chosen to
use an automatic differentiation (AD) approach \cite{rall_1981} so
that we do not need to calculate gradients of $C$ by hand for the two
imaging methods and regularizers.  The use of AD in coherent
diffraction imaging was suggested before powerful parallelized
toolkits were widely available \cite{jurling_josaa_2014}, but it has
since been used for image reconstruction in FFP
\cite{nashed_procedia_2017}, in Bragg and near-field ptychography
\cite{kandel_optexp_2019}, and in NFH and FFP of thick
specimens \cite{du_sciadv_2020}.

\begin{figure}
  \centerline{\includegraphics[width=0.95\textwidth]{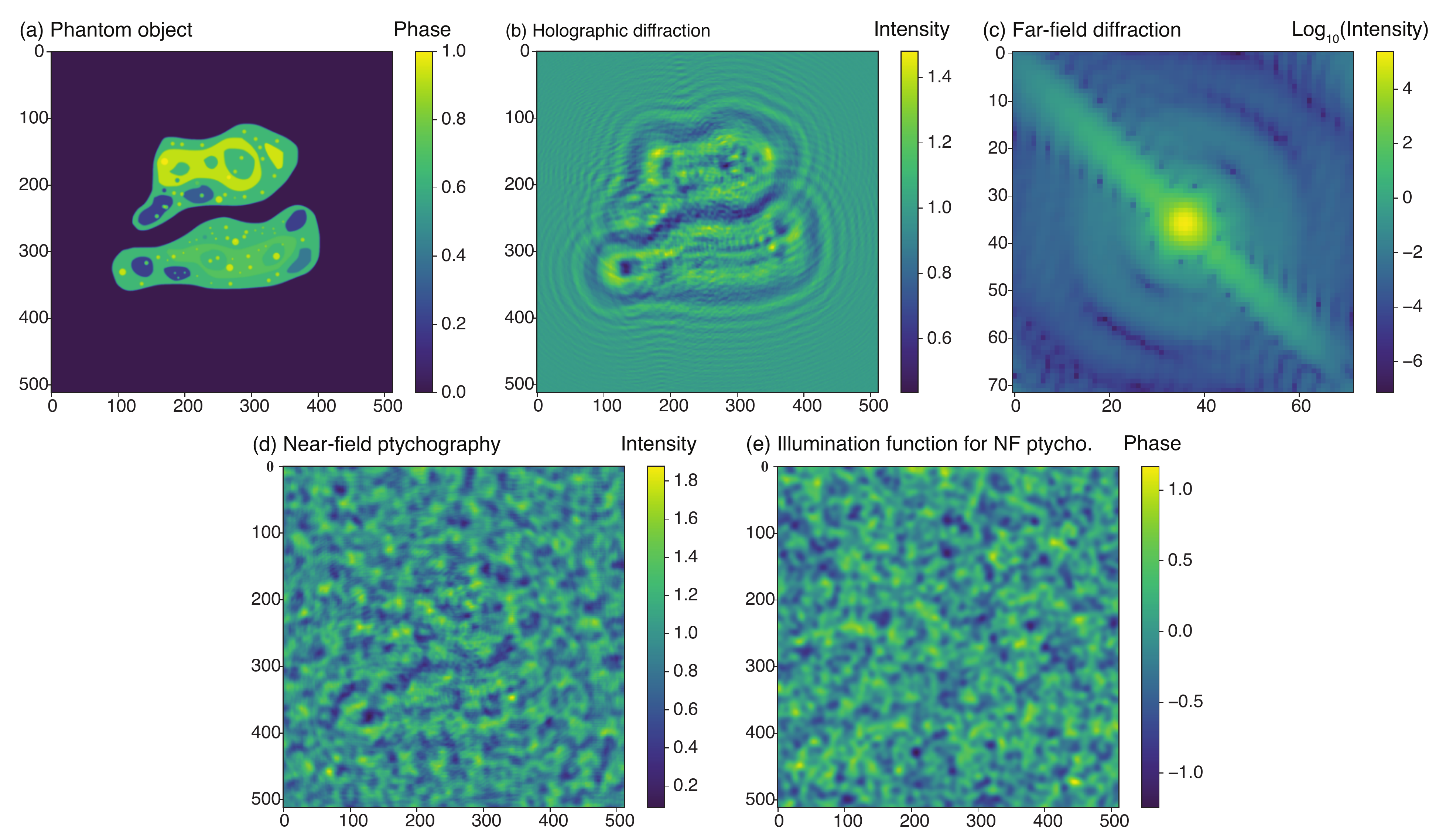}}
  \caption{(a) The $512\times 512$ pixel phantom cell
      object used for our computational experiments. The object is the
      same pure-phase cell phantom used in a prior study
      \cite{hagemann_jac_2017}, so that one can compare directly with
      those results.  The only difference is that we used the complex
      conjugate of the phantom so as to have positive rather than
      negative phase shifts, since x-ray phase is advanced rather than
      retarded in materials \cite{larsson_naturwis_1924}.
      \label{fig:dp} (b) The simulated experimental intensities for
      near-field holography with propagation by a distance
      corresponding to a per-pixel Fresnel number
      (Eq.~\ref{eqn:pixel_fresnel_number}) of $10^{-3}$. (c) One of
      the far-field diffraction patterns at the center of the
      object. In fact, a set of far-field diffraction intensity
      patterns were simulated for a series $k$ of different
      illumination or probe positions across the sample,
    which is the type of dataset one obtains in
      far-field ptychography (FFP). In (d) and (e), we show one of the
      near-field ptychography (NFP) recordings of the object, and the
      phase map of the illumination function used.}
\end{figure}

Our approach is to minimize the cost function $C$ by adjusting the
object function $\vecx$ which contains the complex refractive index of
the sample. For x-ray imaging, we used a 2D grid of the x-ray
refractive index $n(x,y)=1-\delta(x,y)-i\beta(x,y)$ distribution
multiplied by the projection object thickness $t(x,y)$ to yield an
optical modulation of
$\exp\left\{k[i\delta(x,y)-\beta(x,y)]t(x,y)\right\}$ in the sign
convention where forward propagation is $\exp[-ikz]$.  In our case, we
used the same $512\times 512$ pixel pure-phase cell phantom (shown
here in Fig.~\ref{fig:dp}(a)) as was used in prior work
\cite{hagemann_jac_2017}, with the modification of taking its complex
conjugate so that it had positive rather than negative phase values
since x-ray phase is advanced rather than retarded in materials
\cite{larsson_naturwis_1924}.  Within the 19.4\%{} of the pixels that
define the support $S$ of the object, it produces an optical modulation
$\vecx_{0}$ on the incident illumination with a mean phase of
\begin{equation}
  \bar{\varphi}=0.643\mbox{  radians},
  \label{eqn:mean_phase}
\end{equation}
a variance of $\sigma_{\varphi}=0.037$ radians, and a bound of 0 to 1
radians (this object phase contrast is representative of what
one might have in soft x-ray imaging; the contrast is usually lower in
hard x-ray imaging). The cost function $C$ is the mean
squared difference between the modulus of the wave at the detector
plane as predicted by the forward model $f(\vecx, k, d)$ for the
present guess $\vecx$ of the object, and the ``measured'' intensity
$y_{k}$ of
\begin{equation}
  y_{k}=|f(\vecx_{0},k,d)|^{2},
  \label{eqn:y_k}
\end{equation}
where $d$ is the free-space propagation distance $z$ in terms of a
Fresnel number
\begin{equation}
  d=\frac{\Delta^{2}}{\lambda z}
  \label{eqn:pixel_fresnel_number}
\end{equation}
for an object pixel size $\Delta$ (so that far-field diffraction has
$d=0$).  Fresnel propagation $f(\vecx, k, d)$ of the wavefield leaving
the specimen to the detector plane was accomplished via convolution
with a propagator function in the Fourier domain
\cite{goodman_fourier_2017}.  Poisson noise was incorporated in
recorded intensity values $y_{k}$ as will be described below.  We then
had a least-square or LSQ cost function $C_{\text{LSQ}}$ between the
intensities one would expect from the present guess of the object,
versus the measured intensities $y_{k}$, of
\begin{equation}
  C_{\text{LSQ}} = \frac{1}{N_p N_k}\norm{|f(\vecx, k, d)| - \sqrt{y_k}}_2^2.
  \label{eqn:lsq_cost_function}
\end{equation}
where $N_p$ represents the number of pixels in the detector, and $N_k$
represents the number of illumination spots $k$ ($N_{k}=1$ for the
single, full-area illumination in holography). 

The formulation of the cost function in
Eq.~\ref{eqn:lsq_cost_function} is straightforward: by minimizing the
cost function, we update the object function $\vecx$ so that the
Euclidean distance between the diffraction images generated by $\vecx$
and the actual measurements is reduced. A least square (LSQ) cost
function like this is more appropriate for images containing Gaussian
noise which is generally applicable at relatively
high photon fluences \cite{cai_jxst_2017}, but is unable to accurately
account for the shot noise at low photon fluences. When the object is
illuminated by a limited number of photons, the total probability of
observing the entire set of experimental measurement given the object
function $\vecx$ is better described by a Poisson distribution as
\begin{equation}
  p(\vecy|\vecx) = \prod_{i=1}^{N_p N_k}
  \frac{e^{-|f(\vecx, k, d)|_i^2}|f(\vecx, k, d)|_i^{2y_i}}{y_i!}.
  \label{eqn:poisson_likelihood}
\end{equation}
Eq.~\ref{eqn:poisson_likelihood} is also known as the Poisson
likelihood function, and the true object function should be one that
maximizes the likelihood. In practice, the negative logarithm of
Eq.~\ref{eqn:poisson_likelihood} is often taken, so that the
maximization of a serial product can be turned into the more tractable
problem of minimizing a sum. In this way, the Poisson cost function is
written as
\begin{equation}
  C_{\text{Poisson}} = \frac{1}{N_p
    N_k}\sum_{i=1}^{N_pN_k}(|f(\vecx, k, d)|_i^2 - 2y_i\log{|f(\vecx,
    k, d)|_i}).
  \label{eqn:poisson_cost_function}
\end{equation}

In NFP and FFP, the lack of scattering that takes
place when the illuminating probe function is outside the object's
boundary means that it is quite natural for a reconstruction algorithm
to seek solutions for such regions that are empty, even under
conditions of limited illumination.  To add a similar constraint
\textbf{red}{only} to NFH reconstructions, we added to the cost
function of Eq.~\ref{eqn:lsq_cost_function} a regularizer consisting
of a finite support mask $S$.  This yields an update
$\vecx^{\prime}$ to the object of
\begin{eqnarray}
  \vecx^{\prime} &=& \argmin_{\vecx}(C_j) \\
    \nonumber & & \textup{subject to } 
      n_w = 0 \text{ for } n_w \not\in S
      \text{ and } n \ge 0 \text{ for } n \in S\\
    \nonumber & & \text{where } j \in \{\text{LSQ}, \text{Poisson}\}.
    \label{eqn:update_rule}
\end{eqnarray}
A finite
support constraint also suppresses the twin-image in in-line holography
\cite{liu_josaa_1987}. Due to the presence of information redundancy,
FFP and NFP do not need a finite support constraint. 

With the forward model as described above, and the finite support
constraint added to NFH, we were able to obtain
reconstructed images by minimization of the cost function $C$, using
either the LSQ or the Poisson cost function. The partial derivative of $C$ with regards
to the elements of $\vecx$ was calculated using automatic
differentiation (AD) as implemented as a cost function in TensorFlow
\cite{tensorflow_2016}, so that all three imaging types and
both cost function types (LSQ and Poisson) could be treated in
the same way simply by varying the Fresnel number $d$.  The Adam
optimizer \cite{kingma_iclr_2015} in TensorFlow was used to update the
object function using the calculated gradients.

\section{Numerical experiments}

For direct comparison with prior work \cite{hagemann_jac_2017}, we
used the same $512 \times 512$ pixel simulated cell phantom phase
object described above, and the same value of the Fresnel number
(Eq.~\ref{eqn:pixel_fresnel_number}) of $d=10^{-3}$ for NFH. This
corresponds to $z=40.3$ \micron{} with $\Delta=10$ nm pixel size at a
soft x-ray photon energy of 500 eV, or $z=807$ \micron{} at a hard
x-ray photon energy of 10 keV. In the case of NFH, the object was
padded by 256 pixels on each side before optical propagation is
carried out in order to prevent fringe wraparound due to the periodic
array nature of discrete Fourier transforms. The finite support mask
is created by thresholding a low-pass-filtered version of the true
object, so that the mask is about 9 pixels looser than the actual
object boundary. For FFP, we assumed a probe function that was
Gaussian in both magnitude and phase, with a standard deviation of 6
pixels and a phase that varied from 0 to 0.5 radians. The shift
between probe positions was set to 5 pixels so that
there was sufficient probe overlap at low fluence as is required for robust
ptychographic reconstructions \cite{bunk_ultramic_2008}; this is
discussed further in Supplementary Material. This led to a
square scan grid with $66 \times 68$ probe
positions, and for each probe position a $72\times 72$ pixel subset of
the object array was extracted before multiplication with the probe
function and calculation of the resulting $72\times 72$ pixel
diffraction pattern.

For our complementary study on NFP, the setup is
  assumed to be for a point-projection imaging, where a point source
  is used for illumination.  The high spatial resolution of a
  point-projection microscope is achieved by the geometrical
  magnification effect of the spherical wave that the point source
  emits.  As the Fresnel scaling theorem (Appendix B of
  \cite{paganin_2006}) indicates that this geometry is equivalent to
  plane wave illumination with the sample-detector distance scaled by
  a certain factor, we can simulate the image forming process simply
  using a plane wave as the probe function. Since NFP delivers better
  resolution when a diffuser is used to generate a structured
  illumination \cite{stockmar_scirep_2013}, we created our incident
  illumination function as a wavefield with unity magnitude and random
  phase distribution. The phase map was generated by first creating a
  768$\times$768 array of Gaussian-distributed
  per-pixel random phases centered at 0 with
  $\sigma = 0.3$ radian; it was then spatially
  smoothed using a kernel with $\sigma = 5$
  pixels. The phase of the illumination function (cropped to the same
  size as the final diffraction image) is shown in
  Fig.~\ref{fig:dp}(e).  A Fresnel number of $d = 10^{-3}$ between the
  sample and detector, same as the value used for NFH, is used in this
  case. After the sample-modulated wavefield is propagated to the
  detector plane, it was cropped down to a $512\times 512$ to remove
  fringe wrapping at the edges.  Since each diffraction pattern in
  point-projection-based NFP has a much larger effective field-of-view
  (larger than the sample size) compared to FFP, a small number of
  scan spots suffice.  If both the probe function and the object
  function contain $N$ pixels, and so does each diffraction image, 
  then it takes at least 4 diffraction
  patterns to solved both the object and the probe
  \cite{stockmar_scirep_2013}. We therefore followed their choice of
  using 16 diffraction patterns distributed in a $4\times 4$ grid,
  through which the sample translated across the entire
  $512\times 512$ final field-of-view while being fully contained
  inside. This should provide sufficient data for a robust
  reconstruction provided that we use a known probe function.

X-ray microscopes use ionizing radiation, so it is important with many
specimen types to limit the photon fluence $\photonfluence$ (average
number of incident photons that hit each pixel containing the sample)
and consequent radiation dose that the specimen receives. However, one
must supply sufficient fluence in order to successfully image small,
low contrast features. For phase contrast imaging of a non-absorbing,
low-contrast specimen with thickness $t_{f}$ and phase-shifting part
of the refractive index $\delta_{f}$ for feature-containing pixels and
$\delta_{b}$ for background (feature-free) pixels, one can estimate
that the fluence required to obtain an image with a signal to noise
ratio of $\snr$ is given by Eq.~39 of \cite{du_ultramic_2018}, which
we rewrite here as \begin{equation}
  \photonfluence = \frac{\snr^{2}}{2}
  \frac{1}{k^{2}|\delta_{f}-\delta_{b}|^{2}t_{f}^{2}}
  \label{eqn:snr_delta_fluence}
\end{equation}
where $k\equiv 2\pi/\lambda$ is the wavenumber.  Since
$k|\delta_{f}-\delta_{b}|t_{f}$ is the mean phase shift within the
object compared to the object-free region, we can substitute this with
$\bar{\varphi}=0.643$ radians from Eq.~\ref{eqn:mean_phase} and obtain
an estimate that we require a fluence of
\begin{equation}
  \photonfluence=\snr^{2}/[2(\bar{\varphi})^{2}].
  \label{eqn:snr_fluence}
\end{equation}
Given that the variance about the mean phase within the object was
$\sigma_{\varphi}=0.037$ radians, we would expect that a signal to
noise ratio of about $|\bar{\varphi}|/\sigma_{\varphi}=17.4$ would
begin to give very faithful, low noise representations of the true
object, which corresponds to a fluence estimate of
$\photonfluence=350$ photons per pixel (with higher fluences
giving increasing image fidelity).

\begin{figure}
  \centerline{\includegraphics[height=0.6\textheight]{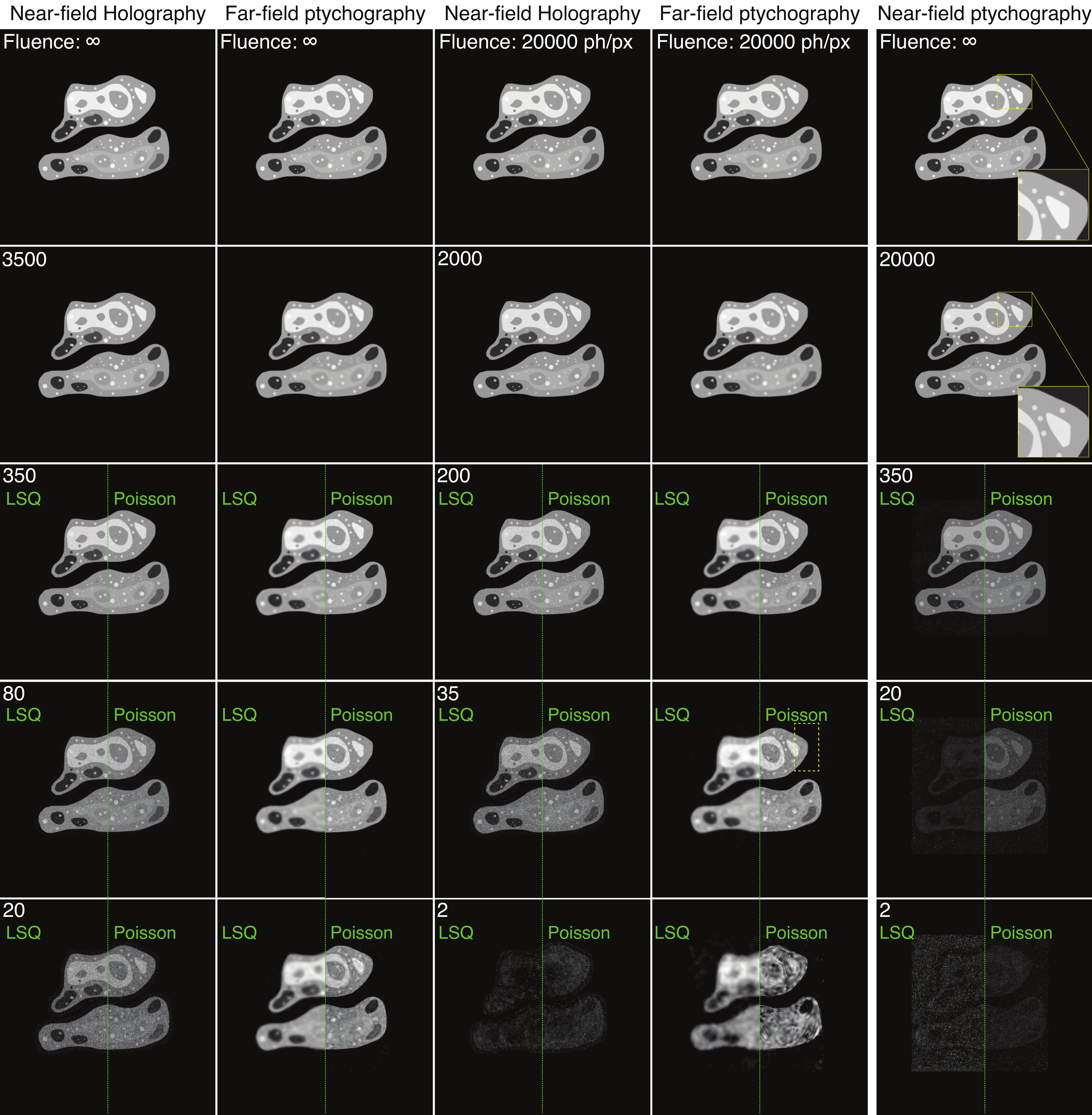}}
  \caption{Reconstructed images of the cell phantom
      shown in Fig.~\ref{fig:dp}(a) obtained for near-field holography
      (NFH), far-field ptychography (FFP), and near-field ptychography
      (NFP) at the photon fluences $\photonfluence$ indicated. For a
      photon fluence higher than 350 photons per pixel, only results
      obtained using the least square (LSQ) cost function are shown;
      for fluences at or below that value,
      we show the reconstructions
      obtained using both the LSQ cost
      function (Eq.~\ref{eqn:lsq_cost_function}) and the Poisson cost
      function (Eq.~\ref{eqn:poisson_cost_function}); these are placed
      side-by-side.  At high photon fluence, both NFH and FFP yield
      high quality images. However, their behaviors differ at low
      fluence. For NFH, the images gain a more salt-and-pepper
      appearance as one would expect from low photon statistics. The
      use of the Poisson noise model does not significantly improve
      the reconstruction quality. In FFP, the decrease in photons
      scattered beyond the illumination probe's numerical aperture at
      low fluence means the images tend more and more towards the
      probe's limit of spatial resolution. While the LSQ cost function
      gives blurry reconstructions at low photon dose, the results
      with the Poisson cost function preserve sharp features even at
      very low photon count, but instead show
      fringe-like artifacts. With NFP, using the
      Poisson cost function at low dose slightly improves the contrast
      in reconstructed images. However, both LSQ and Poisson results
      contain high-frequency artifacts that are eliminated only with
      noise-free diffraction data (see insets).}
  \label{fig:noise_eval_cell}
\end{figure}

We therefore carried out simulations with values of $\photonfluence$
that bracketed a value of $\photonfluence=350$/pixel on an
approximately logarithmic scale. Starting from the noise-free
``recorded'' intensities $y_{k}$ of Eq.~\ref{eqn:y_k}, we incorporated
Poisson noise to $y_{k}$ for a specified total fluence
$\photonfluence$ in photons per pixel on the specimen (to save
computational time, NFP was tested on a subset of the photon fluence
values used for NFH and FFP). Because we expect
$\photonfluence = 350$/pixel to be the nominal dividing line between
``high-dose'' and ``low-dose'' regimes, datasets with $\photonfluence$
beyond that were reconstructed using the LSQ cost function which
approximates photon noise using a quasi-Gaussian model that works well
at high photon fluence. On the other hand, data with $\photonfluence$
below 350/pixel were reconstructed using both the LSQ and the Poisson
cost function. Two separate, independent random noise datasets were
generated for each experiment type, fluence, and loss function type;
reconstructed images from one of these two instances are shown in
Fig.~\ref{fig:noise_eval_cell}. This figure shows that both NFH and
FFP yield high quality reconstructions at high photon fluence. As the
fluence decreases to $\photonfluence=350$/pixel incident photons per
pixel or less, the images begin to show a degradation in quality, but
in different ways. In NFH, the images begin to take on a ``salt and
pepper'' or speckle-like noise appearance as one would expect in a
direct coherent imaging experiment. Switching to the Poisson cost
function does not help significantly with improving the quality. In
FFP at low fluence, one will have relatively few photons scattered
outside the numerical aperture of the probe function, so the image
appears to show a loss of spatial resolution going towards the probe
resolution but with less ``salt and pepper'' noise appearance. At very
low fluences in FFP, there are relatively few photons in the overlap
regions between probe positions. If a sparser scan
  grid was used, one would start to see the scan grid artifacts that
  can arise due to insufficient probe overlap when using the LSQ cost
  function \cite{bunk_ultramic_2008,huang_apl_2017}. The $68\times 66$
  scan grid we used in this case is fine enough to suppress these
  artifacts, but a grid with doubled spot spacing could result in
  obvious grid artifacts, and in that case, the Poisson cost function
  turns out to be a better option (see Supplementary Materials,
  Fig.~S1). The Poisson cost function is also able to give sharper
boundaries of features compared to the LSQ cost function, especially
for $\photonfluence$ below 35/pixel. Nevertheless, results of the
Poisson cost function at relatively high photon fluences incorporate
fringe-like artifacts, such as in the region marked by the yellow
dashed box in the image with $\photonfluence = 35$/pixel. Even when
reconstructing noise-free data, this kind of artifact still exists,
which proves that the Poisson cost function is not always a superior
choice than LSQ and Gaussian cost functions. Another observation
adding to this conclusion is that the Poisson cost function generally
takes more iterations to converge, especially in the case of FFP.

For NFP, using a Poisson cost function improves the
  contrast of the reconstructed images to some extent. However, it was
  observed exclusively in NFP that almost all results obtained from
  noisy data, even with $\photonfluence = 20000$ where NFH and FFP
  yield nearly identical results to the ground truth, contain
  high-frequency artifacts. When the input data are noise-free, then
  NFP is able to reconstruct the image without artifacts, as shown in
  the insets in Fig.~\ref{fig:noise_eval_cell}. The reduced performance at low fluence may be attributed
  to the ambiguity rising from noise-related uncertainty: although
  both a structured illumination and multiple diffraction images are used
  to provide information diversity, the presence of noise makes the
  solution non-unique. Tighter constraint usually leads to a better
  solution, which can be provided either by taking more diffraction
  images so that the uncertainty is reduced by larger sample volume,
  or by using a finite support constraint as in the case of NFH. However,
  a tight support constraint is not always easily determined, and furthermore avoidance of the requirement of a finite support constraint
  is in fact one of the motivations to use NFP.

\begin{figure}
  \centerline{\includegraphics[height=0.55\textheight]{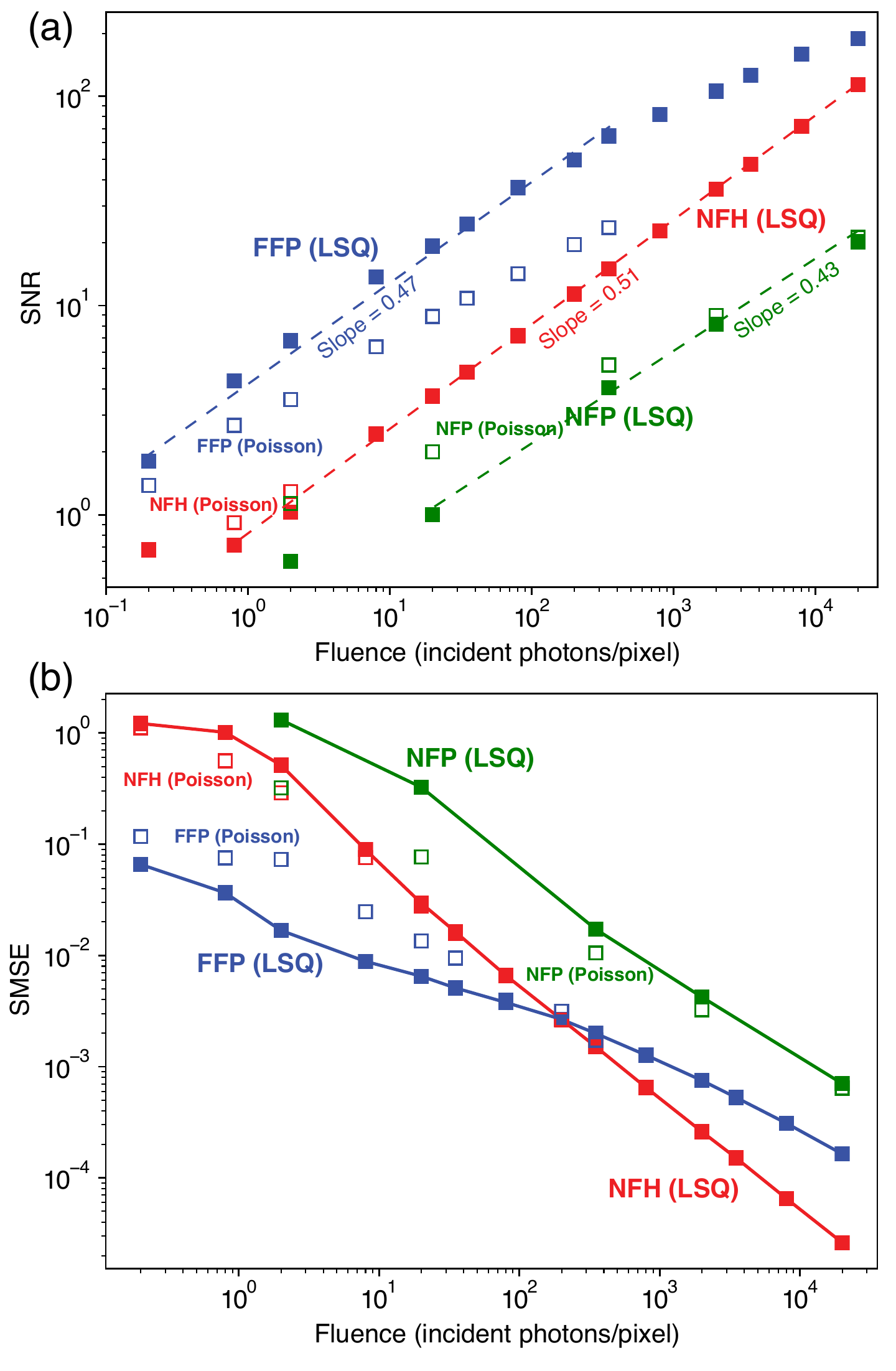}}
  \caption{Whole image metrics of image reconstructed image quality as
    a function of fluence $\photonfluence$.  At top (a) is shown the
    signal-to-noise ratio ($\snr$) as calculated using
    Eq.~\ref{eqn:snr_correlation} for NFH, FFP, and NFP,
    and using either the least-squares (LSQ; solid
      squares) or Poisson cost function (open squares).  The image
  correlation was calculated within the finite support area of the
  object. At each photon fluence $\photonfluence$ and for each cost
  function type, two separate instances of Poisson noise were
  generated and applied to the noise-free dataset. The noisy datasets
  are then independently reconstructed and used for our
  correlation-based SNR calculation. The slope for the least square
  (LSQ) fitting curves is near 0.5 for all three techniques,
  indicating that the $\snr$ increases roughly as
  $\sqrt{\photonfluence}$, as one might expect. At bottom (b) is shown
  the within-support mean squared error (SMSE) of Eq.~\ref{eqn:smse},
  which shows improved performance at low fluences for FFP compared to
  NFH. NFP shows a larger SMSE for all photon fluences tested. }
  \label{fig:snr_smse}
\end{figure}

In order to better quantify the reconstruction quality, we now
consider metrics one can obtain from noisy images. If one has two
images $I_{1}$ and $I_{2}$ of the same object with two different
instances of noise, one can calculate an overall image correlation
coefficient $r$ of \cite{bershad_ieee_1974} \begin{equation}
  r = \frac{\braket{(I_{1}-\braket{I_{1}})\,(I_{2}-\braket{I_{2}})^{\dag}}}
  {\sqrt{\braket{(I_{1}-\braket{I_{1}})^{2}}
      \braket{(I_{2}-\braket{I_{2}})^{2}}}}.
  \label{eqn:r_correlation}
\end{equation}
One can then use this correlation coefficient to calculate
an overall image signal-to-noise ratio \cite{frank_nature_1975} or
$\snr$ of
\begin{equation}
  \snr = \sqrt{\frac{r}{1-r}}
  \label{eqn:snr_correlation}
\end{equation}
where the expression of Eq.~\ref{eqn:snr_correlation} is correct for
intensity images $I_{1}$ and $I_{2}$, as confirmed by the as-expected
scaling of $\snr \propto \sqrt{\photonfluence}$
\cite{huang_optexp_2009}. Although we do not use a
  finite support constraint as part of the NFP and FFP reconstruction
  processes, for comparison with NFH we calculate $r$ and $\snr$ only
  within the finite support region for all three imaging methods
  leading to the result shown in Fig.~\ref{fig:snr_smse}(a).
With the exception of the very lowest fluences in NFH
  and NFP, and NFP fluences above the $\photonfluence=350$ estimate
  given after Eq.~\ref{eqn:snr_fluence} at which one expects to have
  achieved a high fidelity reconstruction of the object, the $\snr$
  from all reconstruction methods show a linear trend on this log-log
  plot with a slope of about 0.5 as expected for
  $\snr \propto \sqrt{\photonfluence}$.  FFP shows the highest overall
  $\snr$, with NFH being second to it, and NFP the lowest. The
  high-frequency and uncorrelated artifacts in NFP results are clearly
  responsible for the method's lower
  SNR. As one compares the results yielded by the two
  types of cost functions, it can be found that while the SNR of NFH
  is slightly enhanced at $\photonfluence = 0.8$ and 2/pixel, the SNR
  of FFP reconstructions with Poisson cost function is actually lower
  than those with LSQ, and the disparity increases at higher
  $\photonfluence$. This observation seems to contradict the visual
  appearance of images in Fig.~\ref{fig:noise_eval_cell}, where
  Poisson reconstructions give sharper feature boundaries under low
  dose conditions. This could be explained by the fact that the method
  of calculating the SNR we have chosen measures the degree of
  correlation between two independently reconstructed images. If the
  images each contain ncorrelated artifacts, the SNR
  is reduced. When using the LSQ cost
  function to reconstruct FFP data, the loss of high-frequency
  information due to photon deficiency results in overall blurriness
  in the reconstructed images. In Poisson reconstructions, however,
  low photon fluence leads to localized fringe artifacts, which are
  heavily dependent on the initial guess. When $\photonfluence$ is
  sufficiently high that LSQ reconstructions are almost noise-free,
  there is still a minor presence of the fringe artifacts in Poisson
  reconstructions. As the initial guess was created by Gaussian noise,
  the positions and amounts of the fringes can vary even for two
  reconstructions corresponding to the same $\photonfluence$. As a
  result, the SNR metric of Eq.~\ref{eqn:snr_correlation} tends to
  interpret the artifacts in FFP reconstructions with Poisson cost
  function as uncorrelated noise.

Since the phantom cell is a pure phase object with a well-defined
support $S$ (which was used in the NFH reconstruction to suppress the
twin image), another whole-image metric we can use is the
within-support mean squared error (SMSE) on the phase of
\begin{equation}
  \mbox{SMSE} =\frac{1}{\sum (n\in S)} \sum_{n \in S} 
  ||\mbox{arg}(\mbox{phantom})-\mbox{arg}(\mbox{reconstruction})||^{2}
  \label{eqn:smse}
\end{equation}
where $n$ is a pixel index.  This is the same $\ell_{2}$-norm metric
defined by Eq.~9 in prior work \cite{hagemann_jac_2017}. Our results for
the SMSE for NFH, FFP, and NFP are shown in
Fig.~\ref{fig:snr_smse}(b).
In \cite{hagemann_jac_2017}, it was
found that NFH gave a higher SMSE at fluences below
about 100 quanta per pixel when compared to far-field CDI, but that
holography then gave a lower SMSE at higher fluences. Here, we have found
a very similar relation between NFH and FFP, with the SMSE cross-over 
also occurring near 100 photons per pixel.
Other than that, we have again found that use of
the Poisson cost function (Eq.~\ref{eqn:poisson_cost_function}) gives slightly
better results than LSQ (Eq.~\ref{eqn:lsq_cost_function}) for NFH and NFP,
but appears to result in larger SMSE for FFP, due to
the more uncorrelated artifacts in FFP's Poisson reconstructions. 

\begin{figure}
  \centerline{\includegraphics[width=0.95\textwidth]{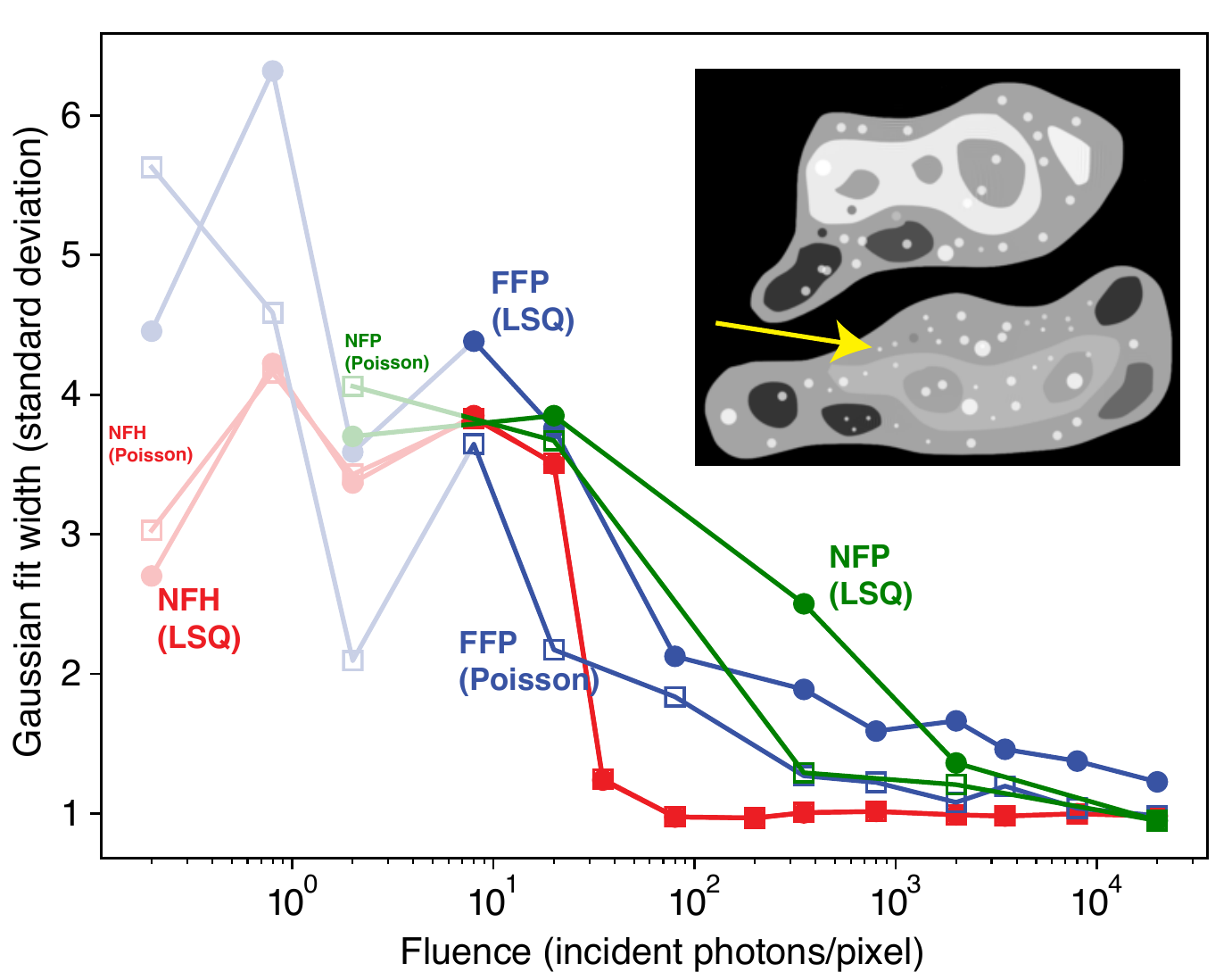}}
  \caption{Standard deviation of the fitted Gaussian 2D profile for the small
  bright-spot-like feature pointed to by the yellow arrow. For photon fluences below 2/pixel,
  the curves are greatly influenced by uncertainties, but meaningful results start
  to appear at higher photon fluences (with outliers for FFP 
  at $\photonfluence = $ 35 and 200/pixel removed). The results agrees with
  the visual appearance of the reconstructions shown in Fig.~\ref{fig:noise_eval_cell},
  where features in FFP reconstructions appear sharper when using the Poisson cost function
  at low photon fluence.}
  \label{fig:sigma}
\end{figure}

Although whole-image SNR measurements show that 
FFP slightly outperforms NFH (and largely outperforms NFP) at low photon
fluence, they also seem to indicate improved results for FFP
when using the LSQ cost function (Eq.~\ref{eqn:lsq_cost_function})
instead of the Poisson cost function
(Eq.~\ref{eqn:poisson_cost_function}) at low fluence, which seems to contradict the
visual appearance of the reconstructed images shown in Fig.~\ref{fig:noise_eval_cell}.
We therefore compared the performance of the NFH, FFP, and NFP
reconstructions for reconstructing a small, bright feature indicated
by a yellow arrow in Fig.~\ref{fig:sigma}. For each
reconstructed image, a Gaussian fit was carried out on this feature
with a 2D symmetric profile, as shown in Fig.~\ref{fig:sigma}.  
An increase in the standard
deviation of the Gaussian fitting function thus measures the blurriness of the
reconstructed image, since a sharper 
feature will have a smaller standard deviation. 
At very low photon fluence, the
overall resolution of the images is low, and the fitted standard deviation may
suffer from significant uncertainty. With $\photonfluence > $ 2/pixel, the results
start to show less fluctuation (outliers for FFP 
at $\photonfluence = $ 35 and 200/pixel have been removed from the plot). 
For FFP, the plot of the Gaussian fit width better indicates the
sharper features brought by Poisson cost function, as the Gaussian 
spread of the feature in 
Poisson-ptychography is smaller than LSQ-ptychography. 
This agrees with visual perception of the results shown in Fig.~\ref{fig:noise_eval_cell}.  

\begin{figure}
  \centerline{\includegraphics[width=0.96\textwidth]{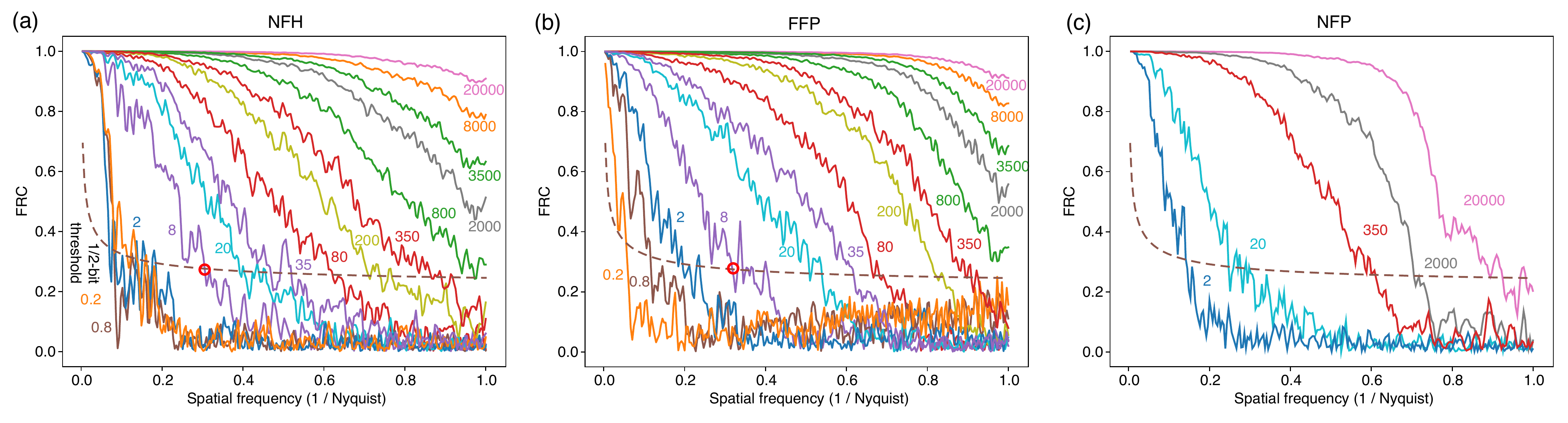}}
  \caption{Fourier ring correlation (FRC) curves for two images
    reconstructed from separate instances of Poisson-noise-included
    simulated datasets, for (a) NFH, (b) FFP, and (c) NFP. 
    Only results obtained using the LSQ
    cost function of Eq.~\ref{eqn:lsq_cost_function}
    are shown. Each curve is labeled with the fluence
    $\photonfluence$ in quanta per pixel.  Also shown on the plot is
    the 1/2-bit threshold curve that is commonly used to define the
    achieved spatial resolution based on the spatial frequency of the
    crossing with the experimental FRC curve \cite{vanheel_jsb_2005},
    as indicated by red circles for a fluence of 8 in (a) and (b). These FRC-crossing normalized spatial frequencies are used
    in Fig.~\ref{fig:frc_crossing}. }
  \label{fig:2d_frc}
\end{figure}

\begin{figure}
  \centerline{\includegraphics[height=0.5\textheight]{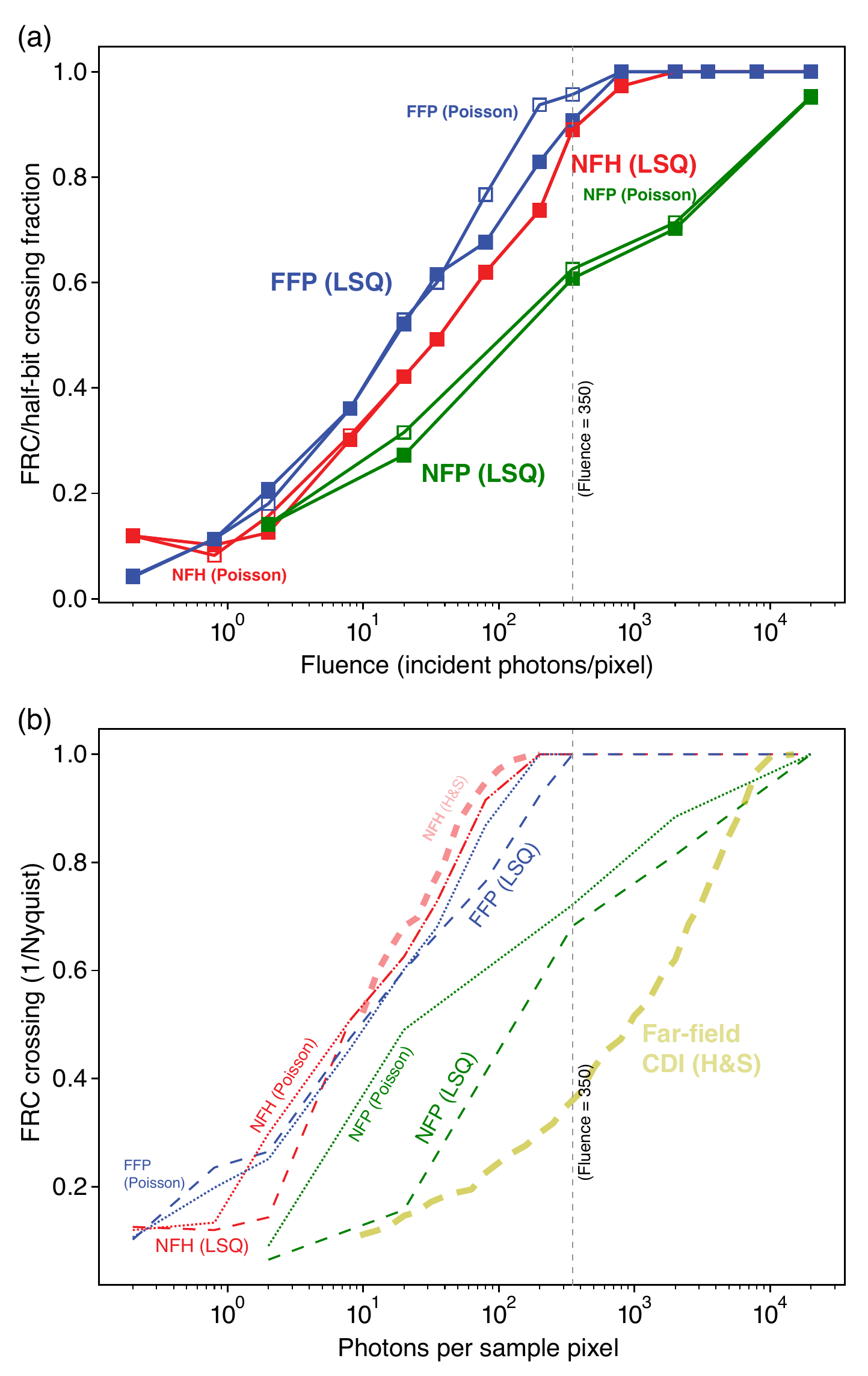}}
  \caption{Values for the crossing between the Fourier ring
    correlation (FRC) curves of Fig.~\ref{fig:2d_frc} and the half-bit
    resolution criterion \cite{vanheel_jsb_2005}, shown as a fraction
    of the Nyquist spatial frequency limit of $1/(2\realpix)$.  In (a),
    this is shown for the FRC analysis between reconstructed images
    obtained from two instances of Poisson noise, as normally
    required.  The curves are not entirely smooth due to the
    sensitivity of the FRC crossing on the exact noise instance of the
    FRC curves shown in Fig.~\ref{fig:2d_frc}, but they show that one
    achieves full spatial resolution with NFH and FFP at fluences near the value of 350
    quanta per pixel (shown with a vertical dashed line) estimated
    after Eq.~\ref{eqn:snr_fluence}.  In prior work
    \cite{hagemann_jac_2017}, the FRC crossing analysis was done by
    comparison of one noise instance with the ``ground truth''
    object of the cell phantom, so (b) shows our results for an
    equivalent ``ground truth'' FRC crossings as dashed lines. 
    Also shown in (b) are the approximate
    results of the previous study \cite{hagemann_jac_2017} labeled with
    ``(H\&S)'' as obtained by tracing of the published
    figure.  (The previous study plotted the FRC crossing as a
    function of $1/(\realpix)$, so we have multiplied the FRC crossing
    fractions by a factor of 2).  As can be seen, our ``ground
    truth analysis'' results and the ``H\&S'' results are reasonably
    consistent for the case of NFH. The previous
    study also considered far-field CDI, where the entire object array
    is illuminated and a finite support constraint is applied during
    iterative reconstruction.}
  \label{fig:frc_crossing}
\end{figure}

Another important metric for evaluating two separate instances of equally noisy
images is to examine the correlation of their Fourier transforms as a
function of radial spatial frequency $\freqr$, leading to the Fourier
shell correlation for 3D images or the Fourier ring correlation (FRC)
for 2D images \cite{saxton_jmic_1982,vanheel_ultramic_1987} given by
\begin{equation}
  \mbox{FRC}_{12}(\freqri) = \frac{\sum_{\freqr \in
      \freqri} F_{1}(r) \cdot F_{2}(r)^{\dag}}{\sqrt{\sum_{\freqr \in
        \freqri} F_{1}^{2}(r) \cdot \sum_{\freqr \in \freqri}
      F_{2}^{2}(r)}}.
  \label{eqn:frc}
\end{equation}
High resolution, low noise images will show strong correlation at high
spatial frequencies, while lower resolution, noisier images will show
poorer correlation at high spatial frequencies.  
It is common to assign a spatial resolution value based on the
crossing of the FRC with a half-bit threshold value
\cite{vanheel_jsb_2005}.
The resulting FRC analysis (plotted only for LSQ results) shown in
Fig.~\ref{fig:2d_frc} indicates that both NFH and
FFP deliver full-resolution images at high photon fluences
with similar information distribution over the spatial frequency 
below the Nyquist limit. On the other hand, NFP largely loses correlativity at
mid-high frequency even at $\photonfluence = 20000$ due to the uncorrelated
artifacts. This figure
also highlights the half-bit resolution FRC crossing point with a red
circle for the case of an incident fluence of 8 quanta per pixel for
each imaging method.  This measure of the spatial resolution as a
fraction of the $1/(2\realpix)$ Nyquist spatial frequency is shown in
Fig.~\ref{fig:frc_crossing}(a), where one can see that both NFH and FFP
approach full resolution at a fluence near the estimate of 350
quanta per pixel found using Eq.~\ref{eqn:snr_fluence}, while NFP
barely reaches full resolution at $\photonfluence = 2\times 10^4$ photons/pixel. 
Because of
the noise fluctuations present in the FRC curves, the FRC/half-bit
crossing fraction may show some variations depending on the particular
instances of data Poisson noise; this explains the non-smooth trend of
the FRC crossing values shown in Fig.~\ref{fig:frc_crossing}(a).


The fraction of the Nyquist limit spatial frequency shown in
Fig.~\ref{fig:frc_crossing}(a) was calculated by FRC analysis from two
separate instances of Poisson noise at each fluence value and each
imaging mode.  However, a prior study has carried out FRC analysis by
comparing a noisy image against the ground-truth image of the
noise-free cell phantom \cite{hagemann_jac_2017}.  We have therefore
calculated this ``ground truth'' FRC crossing value, as well as
tracing the curves shown in Fig.~4(a) of this previous analysis
\cite{hagemann_jac_2017} for both NFH and for
far-field CDI (where the latter involves a single diffraction pattern
from illuminating the entire object array, and the use of a finite
support in iterative phase retrieval).  We  show in
Fig.~\ref{fig:frc_crossing}(b) up to two FRC/half-bit crossing
curves for each experiment/cost function type: the crossing
obtained by comparing one low-fluence image with the ground truth
image (for NFH, FFP, and NFP), 
and the traced values from Fig.~4(a) of the previous analysis
\cite{hagemann_jac_2017} (for NFH and CDI). As can be seen, there is reasonable
agreement betwen our FRC crossing results and those of the previous
analysis \cite{hagemann_jac_2017} for the case of NFH 
with a ground-truth reference, even though the previous
analysis used a slightly different reconstruction algorithm (the
relaxed averaged alternating reflections or RAAR algorithm
\cite{luke_ip_2005}).  In addition, FFP, NFH and NFP all show 
improved performance relative to far-field CDI, which
suffers from well-known difficulties
\cite{miao_prl_2005,thibault_actaa_2006,williams_aca_2007,huang_optexp_2010}.

Overall, the above analyses and discussions suggest similar
performance between far-field ptychography (FFP) and near-field
holography (NFH) over a wide range of fluence, although FFP performs
slightly better in terms of SNR (especially at low photon fluence).
However, it should also be noted that FFP has certain
  extra requirements: it requires a high degree of coherence over the
  entire beam used, while NFH requires high coherence within the
  region of Fresnel fringes from a feature in a specimen but not over
  the entire illumination field.  (As an example, with Fresnel fringes
  extending to 20~\micron{}, one could use a 200~\micron{} wide beam
  with 20~\micron{} coherence width to image a larger field of view
  with higher flux if using a partially coherent source).  FFP also
  requires accurate movement of a probe beam relative to the sample
  (though computational probe position refinement can also help
  correct for errors \cite{guizar_optexp_2008,zhang_optexp_2013}).
Also, all our FFP results shown above were reconstructed with a known
probe function. In reality, it is often the case that the probe needs
to be reconstructed along with the object which is
  straightforward \cite{thibault_science_2008,thibault_ultramic_2009}
  but which also requires additional computation.  Finally, our
  results show poorer performance for near-field ptychography (NFP)
  relative to FFP and NFH, but this may be due in part to the fact
  that we employed a finite support constraint to suppress the twin
  image in NFH, but not in NFP (nor did we use a finite support
  constraint in FFP, since the limited spatial extent of the probe
  function acts in some ways as a per-probe-position finite support
  constraint).  The use of a finite support constraint helps
  tremendously with reconstruction fidelity in NFH, and one could
  expect that it would improve the performance of NFP as well for
  those specimens that do fit within a finite support region.
 
\section{Conclusion}

We have considered a variety of coherent imaging
  methods and how they can perform with varying x-ray fluence.  While
  the brightness (and thus coherent flux) of synchrotron light sources
  has been increasing dramatically (with the next advance being
  provided by diffraction-limited storage rings
  \cite{eriksson_jsr_2014}), radiation dose sets a limit to achievable
  resolution
  \cite{sayre_ultramic_1977,howells_jesrp_2009,du_ultramic_2018}.
  Therefore it is usually desirable to use the lowest fluence
  possible, and instead use increasing coherent flux to image larger
  fields of view with shorter exposure times, or a greater number of
  specimens to give better statistical sampling of a phenomenon.
  
We have used the same automatic-differention-based optimization method
for image reconstruction to compare the performance of near-field
holography, far-field ptychography and near-field ptychography at low
specimen fluence values.  Though this reconstruction algorithm is
slightly different than what was used in a previous study
\cite{hagemann_jac_2017} that compared near-field holography with
single-exposure far-field coherent diffraction imaging (CDI), we have
obtained quite similar results for near-field holography as shown in
Fig.~\ref{fig:frc_crossing}(b), as well as in a comparison of our
Fig.~\ref{fig:snr_smse}(b) with Fig.~4(c) of the previous study.  The
previous study showed that NFH gives greatly superior results compared
to far-field CDI, but far-field CDI is known to be very challenging
due to the experimental difficulty of obtaining an object that has
truly zero scattering outside of a defined region (the finite
support), and due to the sensitivity of the reconstruction to the
correct ``tightness'' of the support and the accuracy of recording the
strong, low-spatial-frequency diffraction signal
\cite{miao_prl_2005,thibault_actaa_2006,williams_aca_2007,huang_optexp_2010}.
Far-field ptychography removes the requirement that the object be
limited to being within a finite support constraint, and if a lens
focus is used to provide the scanned coherent illumination spot the
spreading of the signal in the far-field diffraction pattern helps
reduce the dynamic range demands placed on the detector
\cite{thibault_science_2008}.  In addition, the partitioning of data
recording into a set of distinct regions of the object may provide
some additional information beyond what one obtains when illuminating
the entire object in one exposure, which may be why we observe
slightly improved SNR from FFP relative to NFH in this computational
study.

We conclude that the imaging method used does play some role in the
quality of an image that one can obtain from a given fluence on the
specimen. (We also note that if an optic were to be used to record a
direct image with no reconstruction algorithm required, one would need
to increase the fluence to account for the focusing efficiency of the
optic \cite{huang_optexp_2009} which is often below 20\%{} for the
case of Fresnel zone plates used for x-ray microscopy
\cite{kirz_josa_1974}). However, it is still photon fluence that
dominates the achievable reconstruction, as has long been suggested
based on theoretical analyses
\cite{glaeser_siegel_1975,sayre_ultramic_1977,howells_jesrp_2009,du_ultramic_2018}
and simulation studies \cite{huang_optexp_2009,jahn_aca_2017}.  While
previous studies using NFH suggested that one could obtain images at
reduced radiation dose compared to far-field imaging methods
\cite{bartels_prl_2015}, they did not include a systematic analysis of
resolution versus fluence. Such an analysis was included in a prior
computational study \cite{hagemann_jac_2017}, but it compared NFH with
far-field CDI, rather than with a more robust far-field method like
FFP. When CDI in this comparison is replaced with FFP, we start to
observe that both techniques provide similar spatial resolution at a
wide range of photon fluence, as indicated by our Fourier ring
correlation analysis. By bringing near-field
  ptychography into the comparison, we can
state with more confidence that near-field and
  far-field imaging are generally equivalent in the resolution that
  they can achieve, because information redundancy due to a
  scanning-type acquisition scheme does not necessarily provide an
  advantage, and thus does not really compensate for the resolution of
  FFP. We therefore conclude that the sample can be near, or far;
wherever you are, photon fluence on the specimen sets a fundamental
limit to spatial resolution.

\section*{Acknowledgements}

This research used resources of the Advanced Photon Source and the
Argonne Leadership Computing Facility, which are U.S. Department of
Energy (DOE) Office of Science User Facilities operated for the DOE
Office of Science by Argonne National Laboratory under Contract
No. DE-AC02-06CH11357. We thank the National Institute of Mental
Health, National Institutes of Health, for support under grant R01 
MH115265. We also thank Celine Dion for inspiration for the title.

\bibliographystyle{naturemag}
\bibliography{../../bibtex/mybib}

\end{document}


\maketitle
\renewcommand{\thefigure}{S\arabic{figure}}

\section{Influence of probe spacing in far-field ptychography reconstruction}

\begin{figure}
  \centering
  \includegraphics[width=0.8\textwidth]{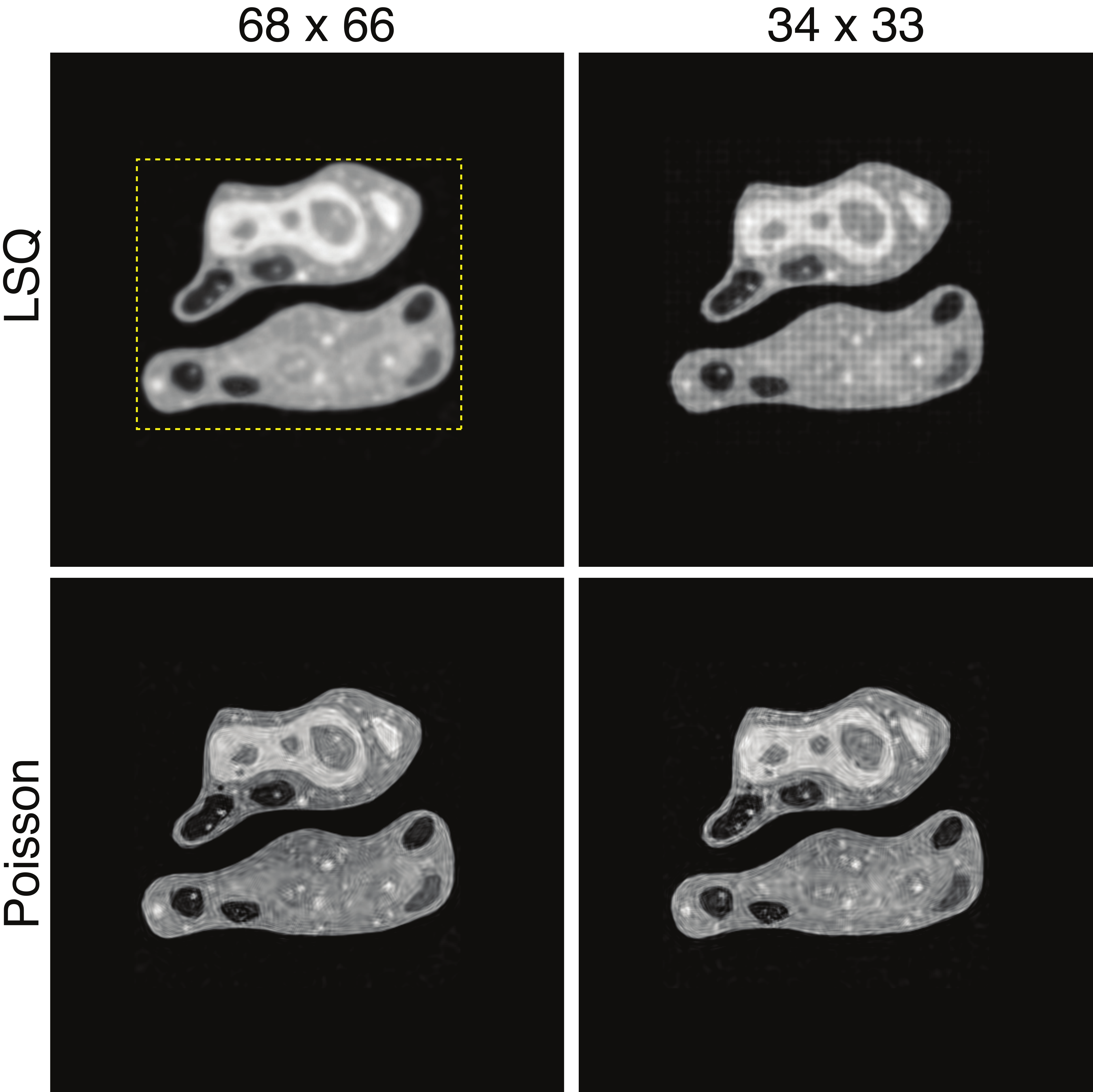}
  \caption{Reconstructions of far-field ptychography (FFP) data with a
    photon fluence of 20 photons per reconstruction
      array pixel using a fine ($68\times 66$ probe positions) and a
      coarse ($34\times 33$ probe positions) scan grid,
      respectively. Reconstructions obtained using the least-squares
    (LSQ) and the Poisson cost function are shown for each scan
    grid. Both grids cover the same area relative to the sample, as
    shown by the yellow dashed box.}
  \label{fig:ptycho_grid}
\end{figure}

When a probe scan grid that is too coarse is used for far-field
ptychography (FFP), the number of photons in the
  ``tails'' of the probe function becomes low enough that there can be
  insufficient overlap between probe positions compared to the
  high-photon-number case. In this case, a grid-like
  artifact will appear in the reconstructed image
\cite{bunk_ultramic_2008,huang_optexp_2014},
especially when using the least-square (LSQ) cost function. The scan
grid of $68\times 66$ probe positions we used to generated all
reconstructions shown in the main text is fine enough to suppress this
artifact at low fluence, but when probe spacing is doubled, the
artifacts become obvious. We demonstrate this effect in
Fig.~\ref{fig:ptycho_grid}, which includes far-field ptychography
(FFP) reconstructions of the dataset generated with a photon fluence
of 20 photons per reconstruction array pixel, using both a fine
($68\times 66$ probe positions) and coarse ($34\times 33$ probe
positions) probe spacing grid.  Both grids covers the same area
indicated by the yellow dashed box, so the coarser grid has twice the
probe spacing in the horizontal and vertical directions.
To keep the number of photons per reconstruction array
  pixel equal (and thus the fluence -- and the dose to the specimen -- equal), we supplied
  1/4 the number of photons per probe position in the fine grid
  relative to the coarse grid.  With the same per-pixel photon dose,
the coarse grid reconstruction using the LSQ cost function, which is
used by popular algorithms like ePIE \cite{maiden_ultramic_2009},
shows obvious grid artifacts.  This grid artifact can
  produce correlations based on artifacts rather than specimen
  features when comparing two independent reconstructions in Fourier
  shell correlation or FSC resolution analysis, and thus can provide a
  biased measure of higher-than-justified resolution at low
  exposures.

When using the Poisson cost function, grid artifacts are
greatly reduced, and the features in the image become
sharper. At the same time, however, another type of artifact
(fringes around fine features, as discussed in the
  main article text) emerges in Poisson reconstructions.

\bibliographystyle{unsrt}
\bibliography{../../bibtex/mybib}